\documentclass[12pt]{article}
\usepackage{scicite}
\usepackage{times}
\usepackage{longtable}
\usepackage{amsmath}
\usepackage{siunitx}
\usepackage{caption}
\usepackage[skip=0.5ex]{subcaption}
\newcounter{lastnote}

\usepackage{graphicx}

\usepackage[modulo]{lineno}
\usepackage{amssymb}
\usepackage{color}
\usepackage{url}

\definecolor{myblue}{RGB}{0,0,160}

\usepackage[colorinlistoftodos,prependcaption,textsize=tiny]{todonotes}

\title{EL meteorites do date the giant planet instability}

\author{Chrysa Avdellidou$^{1,2\ast}$, Marco Delbo'$^{1,2}$, David Nesvorny$^{3}$,\\ Kevin J. Walsh$^{3}$ \& Alessandro Morbidelli$^{1,4}$\\
\\
\normalsize{$^{1}$Laboratoire Lagrange, CNRS, Observatoire de la C\^ote d’Azur, Universit\'e C\^ote d'Azur,}\\
\normalsize{Nice 06304, France}\\
\normalsize{$^{2}$University of Leicester, School of Physics and Astronomy, LE1 7RH, UK}\\
\normalsize{$^{3}$Southwest Research Institute, Boulder, CO, USA} \\
\normalsize{$^{4}$Collège de France, CNRS, PSL Univ., Sorbonne Univ., Paris 75014, France}\\
\\
\normalsize{$^\ast$Corresponding author. E-mail: c.avdellidou@leicester.ac.uk}}

\date{}

\begin{document} 

\baselineskip24pt
\maketitle 


In our recent work \cite{avdellidou2024}, we combined dynamical simulations, meteoritic data and thermal models as well as asteroid observations to argue that the current parent body of the EL meteorites, i.e. asteroid Athor, was implanted into the asteroid belt not earlier than 60~Myr after the beginning of the Solar System and that the most likely capture mechanism was the giant planet orbital instability. This provides a lower limit of 60~Myr to the time of the instability, while an upper limit of 100 Myr is given by the survival of the Patroclus-Menoetious Jupiter Trojan binary system \cite{nesvorny2018}.

In the study ``The link between Athor and EL meteorites does not constrain the timing of the giant planet instability" that appeared in arXiv \cite{izidoro2024arXiv240410828I}, Izidoro and collaborators argue that the implantation of Athor into the asteroid belt does not necessarily require that the giant planet orbital instability occurred at the implantation time. 
They show some simulations where the ejection of a rogue embryo from the terrestrial planet region by Jupiter can cause the implantation of the Athor into the inner main belt. In this case, the capture is due to the secular torquing exerted by the embryo on Athor during the brief high-eccentricity phase that the former acquires just before being ejected.
Izidoro et al. find that the probability that this occurs after 60~Myr is about 1/2 of the probability of the capture of Athor via a late giant planet instability, as we proposed \cite{avdellidou2024}. 

However, terrestrial planet systems with rogue embryos which remain in the system for a long time, such as those illustrated in Izidoro et al. manuscript, tend to be dynamically overexcited in the end. We show this in Fig.~1, where we combined the results of terrestrial planet formation simulations from two different studies \cite{nesvorny2021,woo2024}. Out of a total of 330 simulations only in 50 runs an embryo is ejected from the system after 60~Myr. Of these, only two simulations result in a planetary system similar to the current one, having a normalized angular momentum deficit (AMD)$<$1. Even if we relax the accepted planetary systems to have an AMD$<$1.5 these cases are just five. In other words, only a tenth of the runs can lead to a terrestrial planetary system that is not significantly overexcited. 

The reason for this is that repeated close encounters of terrestrial planets with the embryo stir up their eccentricities and inclinations. Then, subsequent dynamical friction exerted by leftover planetesimals can difficulty damp planets' eccentricity and inclination due to the fast decay of the planetesimal population. 

Moreover, a previous study by Clement et al. \cite{clement2020} showed that the region above the $\nu_6$ secular resonance, where the orbital precession rate is between 24 and 28 arcsec/year, is completely depleted of asteroids, although it is dynamically stable. They convincingly explained this as an effect of the residual migration of Jupiter and Saturn, immediately following the giant planet instability, towards their current orbits near their mutual 2:5 resonance. This implies that no mechanism injected asteroids in this region significantly after that the instability occurred. The capture of Athor during the giant planet instability does not suffer from this problem, because the residual migration of Jupiter and Saturn would have occurred after the capture, clearing the region above the $\nu_6$ resonance as simulated by Clement et al. \cite{clement2020}. 

We argue that, if these constraints (low AMD and no captures above the $\nu_6$ resonance) were taken into account, the probability to implant asteroid Athor after the giant planet orbital instability would result significantly reduced, possibly by an order of magnitude, compared to the result they presented \cite{izidoro2024arXiv240410828I}. 

Thus, in the end, the giant planet instability is still the most likely dynamical process to implant asteroid Athor into the asteroid main belt between 60 and 100~Myr after the beginning of the Solar System.

\bibliographystyle{science}
\bibliography{references}

\begin{figure}
\includegraphics[width=0.9\textwidth]{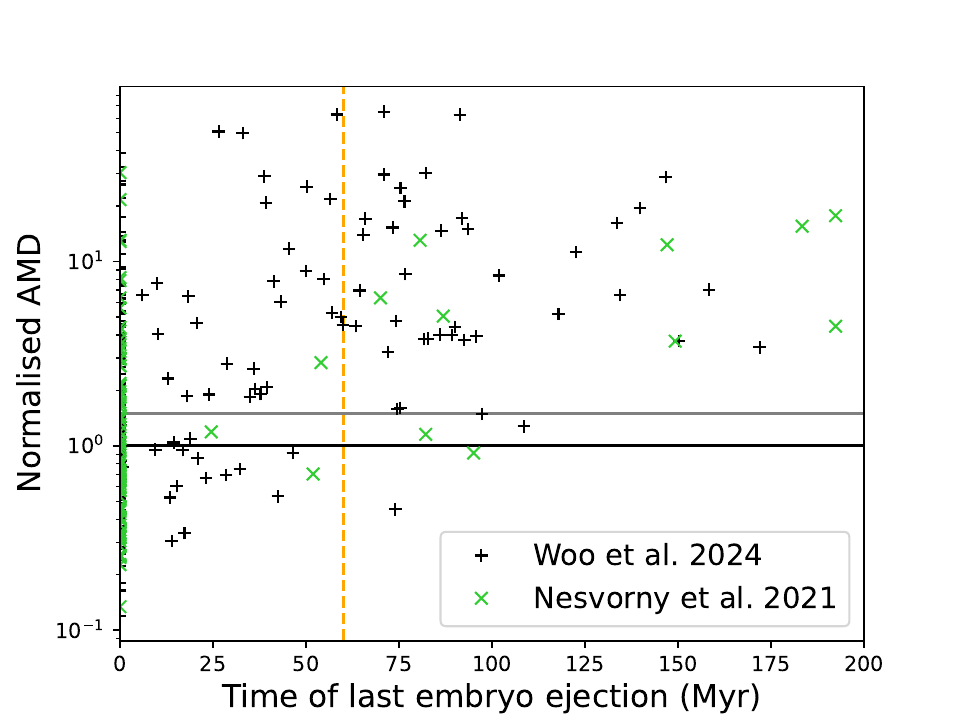} 
\caption{The ejection time of embryos from the terrestrial planet region as a function of the normalized AMD of the simulated terrestrial planet systems (i.e. the ratio between the AMD of the simulated system and of the real terrestrial planets). Black and green markers denote the results from 330 simulations \cite{nesvorny2021,woo2024}. The horizontal black and gray lines show the acceptable range of normalized AMD, while the vertical orange one is drawn at 60~Myr after the formation of the Solar System.}
 \label{FIG:F1}
\end{figure}


\end{document}